# Introduction to Transverse Beam Dynamics

*B.J. Holzer*
CERN, Geneva, Switzerland

**Abstract**
In this chapter we give an introduction to the transverse dynamics of the particles in a synchrotron or storage ring. The emphasis is more on qualitative understanding rather than on mathematical correctness, and a number of simulations are used to demonstrate the physical behaviour of the particles. Starting from the basic principles of how to design the geometry of the ring, we review the transverse motion of the particles, motivate the equation of motion, and show the solutions for typical storage ring elements. Following the usual treatment in the literature, we present a second way to describe the particle beam, using the concept of the emittance of the particle ensemble and the beta function, which reflects the overall focusing properties of the ring. The adiabatic shrinking due to Liouville's theorem is discussed as well as dispersive effects in the most simple case.

## 1    Introduction

The transverse beam dynamics of charged particles in an accelerator describes the movement of single particles under the influence of external transverse bending and focusing fields. It includes the detailed arrangement of the accelerator magnets used (for example, their positions in the machine and their strength) to obtain well-defined and predictable parameters of the stored particle beam. It also describes methods to optimize the single-particle trajectories as well as the dimensions of the beam as an ensemble of many particles. A detailed treatment of this field in full mathematical lucidity and including sophisticated lattice optimizations, such as the right choice of the basic lattice cells, the design of dispersion suppressors or chromaticity compensation schemes, is beyond of the scope of this basic overview. For further reading and for more detailed descriptions, therefore, we would like to refer the reader to more detailed explanations [1–4, 9, 10].

## 2    Geometry of the ring

Magnetic fields are used in general in circular accelerators to provide the bending force and to focus the particle beam. In principle, the use of electrostatic fields would be possible as well, but at high momenta (i.e., if the particle velocity is close to the speed of light), magnetic fields are much more efficient. The force acting on the particles, the Lorentz force, is given by

$$\boldsymbol{F} = q \cdot (\boldsymbol{E} + (\boldsymbol{v} \times \boldsymbol{B}))_.$$

As an example let us consider a magnetic $B$ field of 1 T, which is a conservative size. The resulting Lorentz force acting on a high-energy particle ($v \approx c \approx 3 \times 10^8$ m/s) is

$$F \approx q \cdot 3 \cdot 10^8 \frac{\text{m}}{\text{s}} \cdot 1 \frac{\text{Vs}}{\text{m}^2}$$

$$F \approx q \cdot 300 \frac{\text{MV}}{\text{m}}$$

This corresponds therefore to an equivalent electric field of 300 MV m$^{-1}$, which is far beyond the technical limits, which are set by field breakdown and discharges. In high-energy storage rings, or, more precisely, as soon as the speed of the particles is large enough, magnetic fields are much more efficient in accelerators. It has to be pointed out, however, that for low-energy heavy-ion rings situations might occur where the speed is low enough to make electrostatic fields more efficient.

## 2.1 The ideal circular orbit

We describe the transverse movement of the particles in a coordinate system rotating with the so-called ideal particle. As indicated in Fig. 1, the vertical coordinate, describing the displacement of a particle with respect to the ideal orbit, is called *y*, the corresponding horizontal one *x*, and the coordinate that is pointing into the direction of the longitudinal motion and that is moving with the particles around the ring is called *s*.

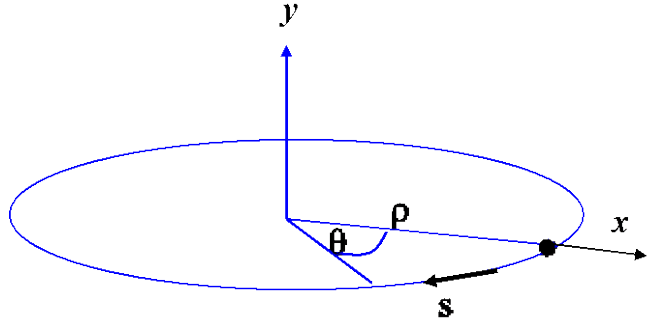

**Fig. 1:** Coordinate system used and orbit of an idealized particle

Assuming ideal conditions and a homogeneous dipole field to provide the bending force, the condition for a circular orbit is given by the equality between the Lorentz force and the centrifugal force. Neglecting any electrostatic field we get

$$q \cdot v \cdot B = \frac{mv^2}{\rho}$$

In a constant transverse magnetic field **B**, a particle will see a constant deflecting force and the trajectory will be part of a circle, whose bending radius $\rho$ is determined by the particle momentum $p = mv$ and the external *B* field:

$$B \cdot \rho = \frac{p}{q}$$

The term *B·ρ* is called the beam rigidity. Inside a dipole magnet, therefore, the bending angle (sketched in Fig. 2) is

$$\alpha = \frac{\int B ds}{B \cdot \rho}. \qquad (1)$$

For the lattice designer, the integrated *B* field along the design orbit of the particles is one the most important parameters, as it is the value that enters Eq. (1) and defines the field strength and the number of such magnets that are installed. By requiring a bending angle of 2π for a full circle, we obtain a condition for the magnetic dipole fields in the ring. Figure 3 shows a photo of a small storage ring [5], where only eight dipole magnets are used to define the design orbit. The magnets are powered symmetrically, and therefore each magnet corresponds to a bending angle $\alpha$ of the beam of exactly 45°. The field strength *B* in this machine is of the order of 1 T.

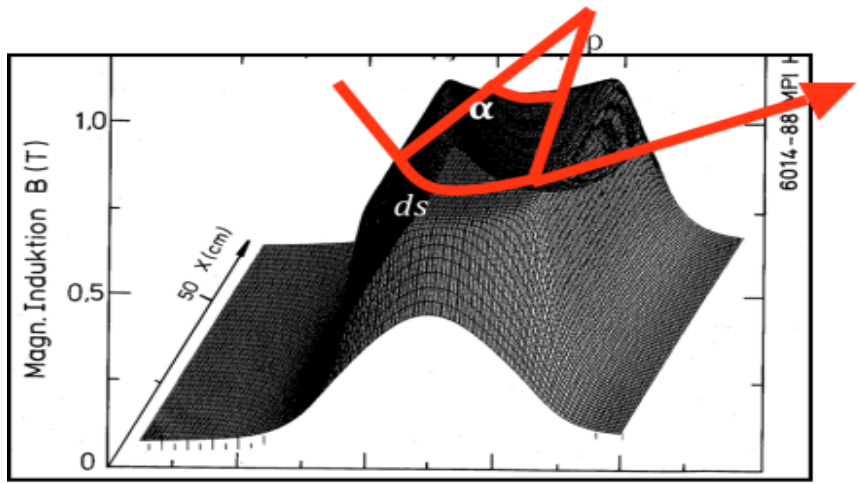

**Fig. 2:** Magnetic *B* field in a storage ring dipole and, schematically, the particle orbit

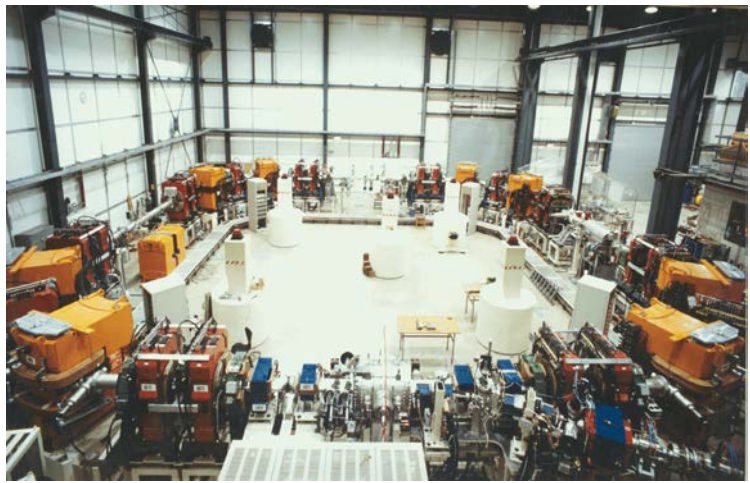

**Fig. 3:** The TSR heavy-ion storage ring at the Max-Planck-Institut in Heidelberg [5]

The lattice and, correspondingly, the beam optics in modern storage rings are usually split into several different characteristic parts. These include arc structures, which are used to guide the particle beam and to establish a regular pattern of focusing elements, leading to a regular, periodic beam dimension. These structures define the geometry of the ring and, as a function of the installed dipole magnets, the maximum energy of the stored particle beam. The arcs are connected by so-called insertions, long lattice sections where the optics is modified to establish the conditions needed for particle injection, to reduce the dispersion function, or to reduce the beam dimensions in order to increase the particle collision rate, for example, where the beam in a collider ring must be prepared for particle collisions.

In the case of the Large Hadron Collider (LHC) at CERN, for a momentum $p = 7000$ GeV/*c*, 1232 dipole magnets are needed in the arc to define the machine geometry, each having a length of 15 m and a *B* field of 8.3 T (Fig. 4). To express this absolutely exactly, the LHC, as any other storage ring, is not really a *ring* but rather a polygon – or, to be even more precise, a 1232-gon.

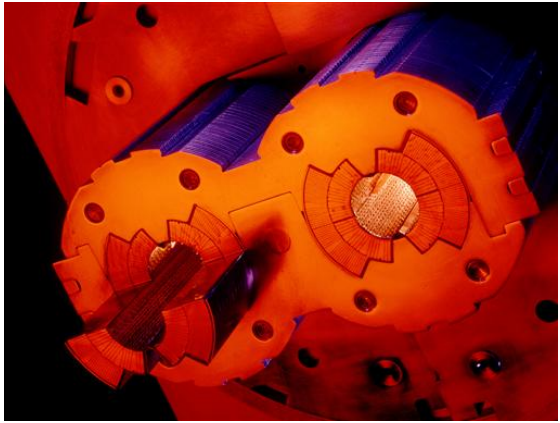

**Fig. 4:** The s.c twin-aperture dipole magnets of LHC

In general, we will try to cover a maximum part of the circumference by the main bending magnets, as they define the highest energy of the machine. For technical reasons, as a general rule about two-thirds of the circumference of the machine can be used to install the dipoles.

Requiring an overall bending angle of $2\pi$ for the complete dipole bending strength and approximating the integral of the $B$ field by the sum over the magnets, we can calculate the required $B$ field. In the case of the LHC we get

$$\oint B dl \approx N\, l\, B = 2\pi\, p/e$$

$$B \approx \frac{2\pi \cdot 7000 \cdot 10^9 \text{ eV}}{1232 \cdot 15\text{m} \cdot 3 \cdot 10^8 \frac{\text{m}}{\text{s}} \cdot e} = 8.3 \text{ T}$$

A part of the LHC tunnel is shown in Fig. 5, where the shape of the ring is hardly visible due to the large bending radius of $\rho$ = 2.5 km.

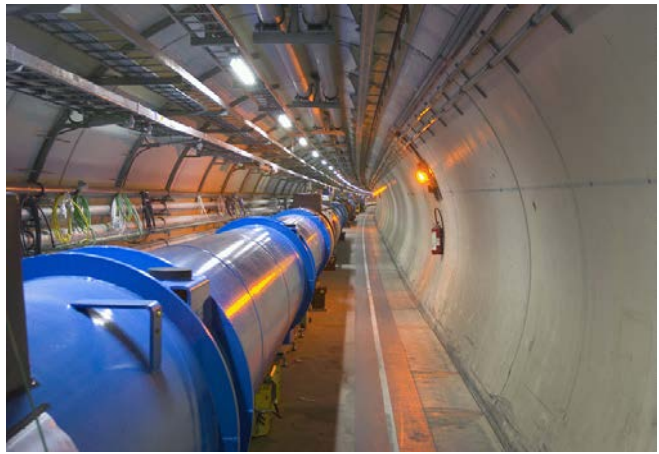

**Fig. 5:** The LHC tunnel with the dipole cryostats. The normalized bending force, represented by the radius of $\rho$ = 2.5 km, is hardly visible.

## 3 Quadrupole magnets and equation of motion

Once the geometry and specification of the arc have been determined and the layout of the bending magnets has been done, the next step is to worry about the focusing properties of the machine. In general, we have to keep more than $10^{12}$ particles in the machine, distributed over a number of bunches, and these particles have to be focused to keep their trajectories close to the design orbit.

Gradient fields generated by quadrupole lenses are used to do this job. These lenses generate a magnetic field that increases linearly as a function of the distance from the magnet centre:

$$B_y = -g \cdot x, \quad B_x = -g \cdot y$$

Here, $x$ and $y$ refer to the horizontal and vertical planes and the parameter $g$ is called the gradient of the magnetic field. It is customary to normalize the magnetic fields to the momentum of the particles. In the case of dipole fields, we obtain from Eq. (1)

$$\alpha = \frac{\int B ds}{B \cdot \rho} = \frac{1}{\rho} \cdot L_{eff}$$

where $L_{\text{eff}}$ is the so-called effective length of the magnet. The term $1/\rho$ is the normalized bending strength of the dipole. In the same way, the field of the quadrupole lenses is normalized to $B\rho$. The strength $k$ is defined by,

$$k = \frac{g}{B \cdot \rho}$$

and the focal length of the quadrupole is given by

$$f = \frac{1}{k \cdot l}$$

As a rule of thumb, we get the normalized gradient $k$ as the ratio between gradient $g$ expressed in T m$^{-1}$ and the particle momentum $p$ in units of GeV/$c$ and multiplying by 0.3 as an approximation of the speed of light:

$$k = 0.3 \frac{g(\frac{T}{m})}{p(\frac{GeV}{c})}$$

It is worth emphasizing that the gradient $g$ will describe the focusing property of the quadrupole in both transverse planes. From Maxwell's equation,

$$\nabla \times \boldsymbol{B} = \cancel{j} + \cancel{\frac{\partial E}{\partial t}} = 0$$

because at the position of the beam (i.e., inside the quadrupole bore) the density $j$ of the magnet current is zero and no explicit changing electric field exists. We conclude that

$$\frac{\partial B_y}{\partial x} = \frac{\partial B_x}{\partial y}$$

To derive the equation of motion of the particles under the influence of the fields described above, we refer to the linear approximation. In the case of the fields this means that only constant terms or terms linearly dependent on $x$ (or $y$) will be taken into account. Mathematically speaking, for a general Taylor expansion, all higher-order terms will be neglected:

$$\frac{\boldsymbol{B}(x)}{p/e} = \frac{1}{\rho} + k\,x + \cancel{\frac{1}{2!} m\, x^2} + \cancel{\frac{1}{3!} n\, x^3} + \dots$$

This approximation is more serious than it might look like. Nonlinear terms would lead to an equation of motion that cannot be solved analytically any more. For the storage ring design, the consequence is to optimize the magnets and suppress higher multipoles in the dipole and quadrupole magnets as well as possible. Moreover, in general in modern accelerators, the magnets will be split and optimized according to their job: dipoles for bending, and quadrupoles for focusing. Multipoles like sextupole and octupole magnets will be introduced – if needed – carefully for higher-order corrections only.

In Fig. 3 we saw already an example of such a separate function storage ring [5] with eight dipole magnets and 4 × 5 quadrupole lenses for beam focusing.

### 3.1 Equation of motion

For the derivation of the equation of motion, we start with a general expression for the radial acceleration as known from classical mechanics:

$$a_r = \frac{d^2\rho}{dt^2} - \rho\left(\frac{d\theta}{dt}\right)^2$$

The first term refers to an explicit change of the bending radius, and the second one to the centrifugal acceleration. Referring to our coordinate system, and replacing for the general case the ideal radius $\rho$ by $\rho + x$ (Fig. 6), we obtain for the balance between radial force and the counter-acting Lorentz force the relation

$$F = m\frac{d^2}{dt^2}(x+\rho) - \frac{mv^2}{x+\rho} = e B_y v$$

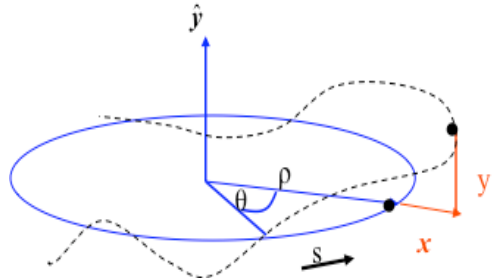

**Fig. 6:** Ideal circular orbit and real particle trajectory with its transverse coordinates $x$ and $y$

On the right-hand side of the equation, we take only linear terms into account,

$$B_y = B_0 + x\frac{\partial B_y}{\partial x}$$

and for convenience we replace the independent variable $t$ by the coordinate $s$,

$$x' = \frac{dx}{ds} = \frac{dx}{dt}\frac{dt}{ds}$$

to obtain an expression for the particle trajectories under the influence of the focusing properties of the quadrupole and dipole fields in the ring, described by a differential equation. This equation is derived in its full beauty elsewhere [6], so we shall just state it here:

$$x'' + Kx = 0. \tag{2}$$

Here, $x$ describes the horizontal coordinate of the particle with respect to the design orbit; the derivative is taken with respect to the orbit coordinate $s$, as usual in linear beam optics; and the parameter $K$ combines the focusing strength $k$ of the quadrupole and the weak focusing term $1/\rho^2$ of the dipole field. (Note that a negative value of $k$ means a horizontal focusing magnet.) The value of $K$ is given by

$$K = -k + 1/\rho^2.$$

In the vertical plane, in general, the term $1/\rho^2$ is missing, as in most (but not all) accelerators the design orbit is in the horizontal plane and no vertical bending strength is present. At the same time, the sign of the gradient changes due to the geometry of the quadrupole field lines, $k \to -k$. So, in the vertical plane, we have

$$K = k.$$

## 3.2 Single-particle trajectories

The differential equation (2) describes the transverse motion of a particle with respect to the design orbit. This equation can be solved in a linear approximation, and the solutions for the horizontal and vertical planes are independent of each other in the sense that the motions in the two transverse planes are uncoupled.

If the focusing parameter $K$ is constant, which means that we are referring to a position inside a magnet where the field is constant along the orbit, the general solution for the position and angle of the trajectory can be derived as a function of the initial conditions $x_0$ and $x_0'$. In the case of a focusing lens, we obtain

$$x(s) = x_0 * \cos(\sqrt{K}*s) + \frac{x_0'}{\sqrt{K}} * \sin(\sqrt{K}*s),$$

$$x'(s) = -x_0 * \sqrt{K} * \sin(\sqrt{K}*s) + x_0' * \cos(\sqrt{K}*s),$$

or, written in a more convenient matrix form,

$$\begin{pmatrix} x \\ x' \end{pmatrix}_s = M \cdot \begin{pmatrix} x \\ x' \end{pmatrix}_0.$$

Given the particle amplitude and angle in front of the lattice element, $x_0$ and $x_0'$, we obtain their values after the element by a simple matrix multiplication. The matrix $M$ depends on the properties of the magnet, and we obtain the following expressions for three typical lattice elements. Also shown schematically are the situations for the three cases described.

• Focusing quadrupole:

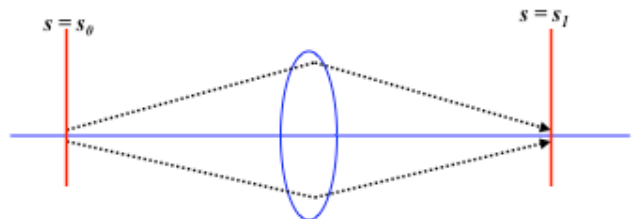

$$M_{QF} = \begin{pmatrix} \cos(\sqrt{K}l) & \frac{1}{\sqrt{K}}\sin(\sqrt{K}l) \\ -\sqrt{K}\sin(\sqrt{K}l) & \cos(\sqrt{K}l) \end{pmatrix}, \tag{3a}$$

- Defocusing quadrupole:

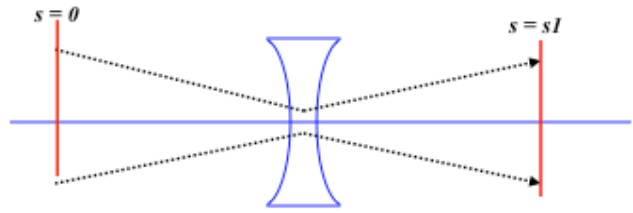

$$M_{QD} = \begin{pmatrix} \cosh(\sqrt{K}l) & \frac{1}{\sqrt{K}}\sinh(\sqrt{K}l) \\ \sqrt{K}\sinh(\sqrt{K}l) & \cosh(\sqrt{K}l) \end{pmatrix}, \quad (3b)$$

- Drift space:

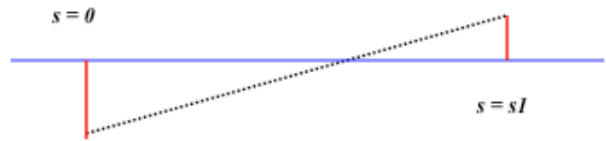

$$M_{\text{drift}} = \begin{pmatrix} 1 & \ell \\ 0 & 1 \end{pmatrix} \quad (3c)$$

### 3.3 Transformation through a system of lattice elements

Given the matrix description for the lattice elements, it is evident that, to obtain the particle amplitude at a certain position in the storage ring, we can either start at an initial point and transform the vector $(x, x')$ step-by-step through the machine, or – mathematically equivalent, but much more convenient – determine the product matrix between the two points of interest and apply it to the initial coordinates $(x, x')_0$. As an example, we refer to Fig. 7. For simplicity, we consider a storage ring built only out of focusing and defocusing quadrupoles and dipole magnets in between.

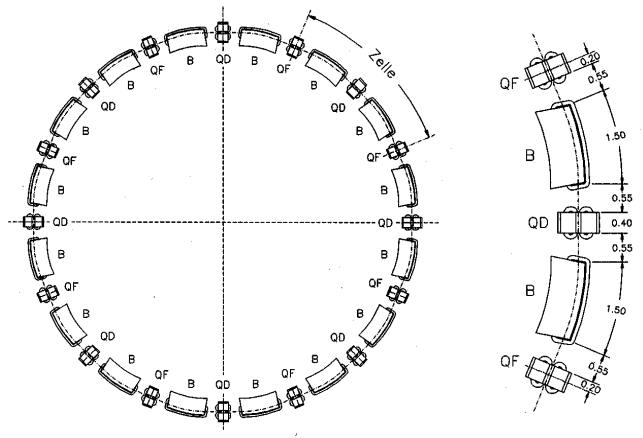

**Fig. 7**: Ideal storage ring consisting of focusing and defocusing quadrupoles and dipole magnets for bending (Courtesy of K. Wille [7]).

Starting in front of a focusing magnet, a typical part of this structure, expressed in matrix form, would be e.g.

$$M_{total} = M_{QF} * M_D * M_{QD} * M_{Bend} * M_{D*}....$$

The particle coordinates at point $s_2$ can be obtained by repetitive multiplication of the corresponding single element matrices, or by applying the product matrix above:

$$\begin{pmatrix} x \\ x' \end{pmatrix}_{s2} = M(s_2, s_1) * \begin{pmatrix} x \\ x' \end{pmatrix}_{s1}$$

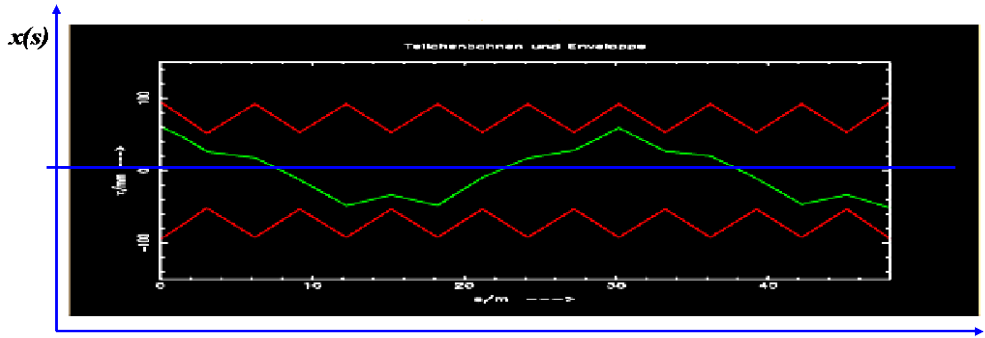

**Fig. 8**: Single-particle trajectory in a storage ring (green line). In red, the pattern of the beta function is shown, which can be interpreted as the maximum beam size or aperture needed in the example.

Using reasonable values for $1/\rho$ and $k$, we obtain the picture shown in Fig. 8 for the particle trajectory. The particle movement – looking zig-zag in the simplified model used here – represents in each single element the solution of a harmonic oscillator, each having eventually a different restoring force, still however leading to an overall focusing effect around the storage ring. The overall number of oscillations, or better the corresponding oscillation frequency of this transverse motion, is of major importance for beam stability reasons and represents the eigenfrequency of the motion, the so-called 'tune'. In the example of Fig. 8 we get $Q \approx 1.3$.

## 4  The Twiss parameters $\alpha$, $\beta$ and $\gamma$

Following the single-particle trajectory as shown in Fig. 8, the natural question will be how the trajectory of the second turn will look; and then the third one; and so on. Reproducing the calculations that lead to the orbit of the first turn will result in a large number of single-particle trajectories that will overlap somehow and form the beam envelope. Figure 9 shows the result for 50 turns. Clearly, as soon as we talk about many turns, or many particles, the use of the single-trajectory approach is quite limited and we need a description of the beam as an ensemble of many particles.

Fortunately, in the case of periodic conditions in the accelerator, there is another way to describe the particle trajectories that, in many cases, is more convenient than the above-mentioned formalism, which is valid within a single element. It is important to note that, in a circular accelerator, the focusing elements are necessarily periodic in the orbit coordinate $s$ after one revolution. Furthermore, storage ring lattices have in most cases an inner periodicity: they are often constructed, at least partly, from sequences in which identical magnetic cells, the lattice cells, are repeated several times in the ring and lead to periodically repeated focusing properties.

In this case, the equation of motion has to be written now in a slightly different form:

$$x''(s) - k(s)x(s) = 0$$

Unlike the treatment above, the focusing parameters – or the restoring forces – are no longer constant, but are functions of the coordinate *s*. However, they are periodic in the sense that, at least after one full turn, they repeat each other, $k(s + L) = k(s)$, leading to the special so-called Hill differential equation. Following Floquet's theorem, the solution of this equation can be written in its general form as [2]

$$x(s) = \sqrt{\varepsilon} \cdot \sqrt{\beta} \cdot \cos(\psi(s) + \phi) \tag{4}$$

$$x'(s) = \frac{-\sqrt{\varepsilon}}{\sqrt{\beta(s)}} \cdot \sin(\psi(s) + \phi) + \alpha(s) \cdot \cos(\psi(s) + \phi)$$

The position and angle of the transverse oscillation of a particle at a point *s* are given by the value of a special *β*-function at that location, and *ε* and *δ* are constants of the particular trajectory. The *β*-function depends in a quite sophisticated manner on the overall focusing properties of the storage ring. It cannot be calculated directly in an analytical approach, but it has to be either determined numerically or deduced from properties of the single-element matrices described above (see e.g. [8]). In any case, like the lattice itself, it has to fulfil the periodicity condition

$$\beta(s + L) = \beta(s)$$

Inserting the solution (4) into the Hill equation and rearranging slightly, we get

$$\psi(s) = \int_0^s \frac{ds}{\beta(s)} \tag{5}$$

which describes the phase advance of the oscillation. It should be emphasized that *ψ* depends on the amplitude of the particle oscillation. At locations where *β* reaches large values, i.e., the beam has a large transverse dimension, the corresponding phase advance will be small; and vice versa, at locations where we create a small *β* in the lattice, we will obtain a large phase advance. In the context of Fig. 8 we introduced the tune as the number of oscillations per turn, which is nothing other than the overall phase advance of the transverse oscillation per revolution in units of 2π. So by integrating Eq. (5) around the ring, we get the expression

$$Q_y = \frac{1}{2\pi} \oint \frac{ds}{\beta(s)}$$

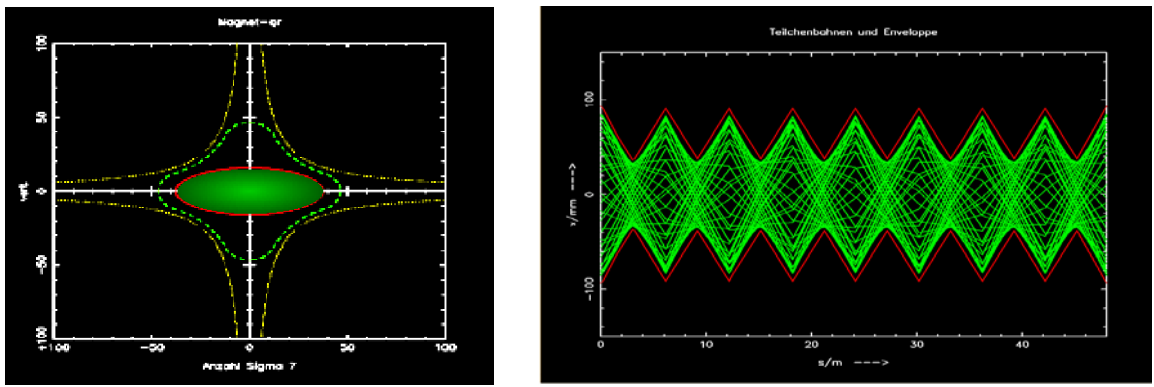

**Fig. 9**: Beam as an ensemble of many single-particle trajectories. Left: transverse shape of a typical particle beam. Right: overlapping of many particle trajectories.

The practical significance of the *β*-function is shown in Fig. 9. The left part shows schematically the transverse shape of the beam inside the vacuum chamber (red) and the hyperbolic pole shoe profile of the quadrupole lens. On the right-hand side the single-particle trajectories are shown. The envelope of the overlapping trajectories is given by $\hat{x} = \sqrt{\varepsilon \beta(s)}$ and is used to define the beam size in the sense of a Gaussian density distribution.

The integration constant $\varepsilon$ has a clearly defined physical interpretation. Given the solution of Hill's equation

$$x(s) = \sqrt{\varepsilon} \sqrt{\beta(s)} \cos(\psi(s) + \phi) \tag{6}$$

and its derivative

$$x'(s) = -\frac{\sqrt{\varepsilon}}{\sqrt{\beta(s)}} \{\alpha(s) \cos(\psi(s) + \phi) + \sin(\psi(s) + \phi)\} \tag{7}$$

we can transform Eq. (6) to

$$\cos(\psi(s) + \phi) = \frac{x(s)}{\sqrt{\varepsilon} \sqrt{\beta(s)}}$$

and insert the expression into Eq. (7) to get an expression for the integration constant $\varepsilon$:

$$\varepsilon = \gamma(s) x^2(s) + 2\alpha(s) x(s) x'(s) + \beta(s) x'^2(s) \tag{8}$$

Here we follow the usual convention in the literature and introduce the two parameters

$$\alpha(s) = \frac{-1}{2} \beta'(s)$$

$$\gamma(s) = \frac{1 + \alpha(s)^2}{\beta(s)} \tag{9}$$

We obtain for $\varepsilon$ a parametric representation of an ellipse in $x$–$x'$ 'phase' space, which according to Liouville's theorem is a constant of motion, as long as conservative forces are considered. The mathematical integration constant thus gains physical meaning.

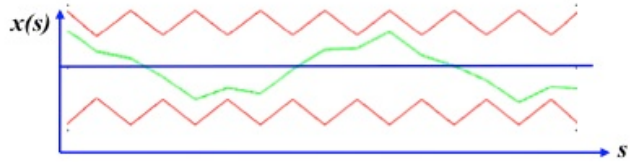

**Fig. 10:** A single-particle trajectory …

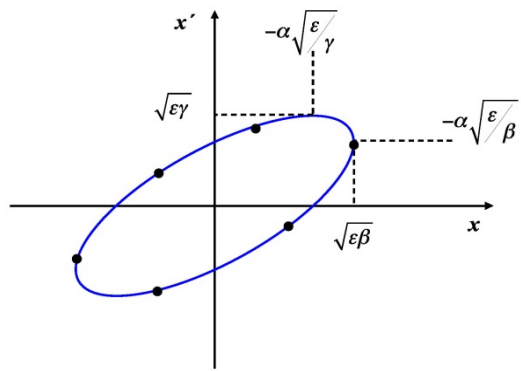

**Fig. 11:** … and its phase space coordinates at a position $s$ turn by turn

Referring again to the single-particle trajectory, discussed above (Fig. 10), but now plotting at a given position $s$ in the ring the coordinates $x$ and $x'$ turn-by-turn, we get the picture shown in Fig. 11. Plotted in phase space the particle will follow the shape of an ellipse whose shape and orientation are defined by the optical parameters at the reference position *s* in the ring. Each point in Fig. 11 represents the coordinates for a certain turn at that position in the ring, and the particle will have performed from one turn to the next a number of revolutions in phase space that corresponds to its tune. It has to be emphasized that, as long as conservative forces are considered (i.e., no interaction between the particles in a bunch, no collisions with remaining gas molecules, no radiation effects, etc.), the size of the ellipse in $x$–$x'$ space is constant and can be considered as a quality factor of the single particle. Large areas in $x$–$x'$ space (where, to be exact, the area is given as $A = \pi\varepsilon$) mean large amplitudes and angles of the transverse particle motion, and we would consider it as a bad particle 'quality'.

To discuss the dependence of the shape of the ellipse on the beam optics in a bit more detail, we replace the two parameters $\alpha$ and $\gamma$ by their corresponding expressions in Eq. (9),

$$\varepsilon = \frac{x^2}{\beta} + \frac{\alpha^2 x^2}{\beta} + 2\alpha \cdot xx' + \beta \cdot x'^2$$

solve for $x'$

$$x'_{1,2} = \frac{-\alpha \cdot x \pm \sqrt{\varepsilon\beta - x^2}}{\beta}$$

and determine the maximum values of the particle angle $\hat{x}'$ via

$$\frac{dx'}{dx} = 0$$

to obtain for the maximum angle $x'$ and the position $x$ at that point:

$$\hat{x}' = \sqrt{\varepsilon\gamma}, \qquad x = \pm\alpha\sqrt{\varepsilon/\gamma}$$

For completeness and to demonstrate the conservation of phase space area, we plot in Fig. 12 the emittance invariants of four particles in phase space. For the position *s* this time the centre of a focusing quadrupole has been chosen, where the orientation of the ellipses is flat, due to the fact that at this location the beam will have its largest size, i.e., $\alpha = 0$. Turn by turn the coordinates of the four particles will lie on the corresponding ellipses and never cross each other.

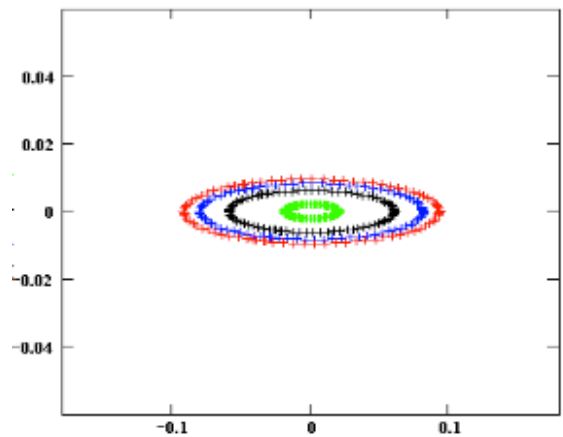

**Fig. 12:** Phase space ellipses at the location of a focusing quadrupole for four particles

## 4.1 The emittance of a particle ensemble

The ensemble of many single particles, as shown in Fig. 13, forms a pattern of overlapping trajectories that in the end we will observe as transverse intensity (or charge) distribution and that we will use to define the beam size. Along the storage ring coordinate *s*, its transverse size, in the sense of the maximum amplitude of a trajectory that will be observed at a given location, is defined by the *β*-function.

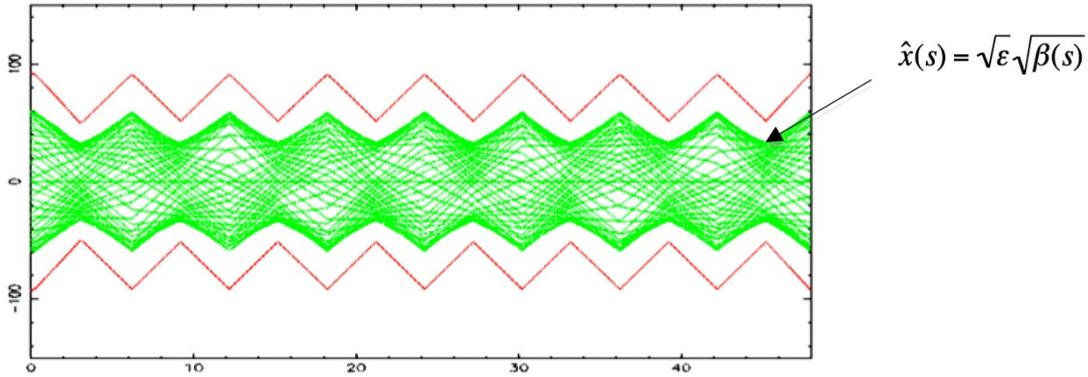

**Fig. 13**: The overlapping trajectories of many single particles define the beam cross-section as defined by ε and β.

As the general solution of the equation of motion is given by

$$x(s) = \sqrt{\varepsilon}\sqrt{\beta(s)} \cdot \cos(\Psi(s) + \phi)$$

the maximum amplitude, or beam size, is obtained from

$$\hat{x}(s) = \sqrt{\varepsilon}\sqrt{\beta(s)}$$

and depends via the *β*-function on the focusing properties of the lattice and via *ε* on the quality of the particle ensemble.

In many cases, the transverse particle density of the particles follows a Gaussian distribution. Referring to a particle within this distribution that is situated at one standard deviation *σ*, the *ε* parameter (sometimes called the Courant–Snyder invariant) of this particle can be used as representative of the complete beam. In this sense we talk about a beam emittance and thus about a general quality factor of the whole particle ensemble. Referring to the example of the LHC, for example, we have, in the arc of the storage ring at flat-top energy, the following values:

$$\beta = 180 \text{ m}$$
$$\varepsilon = 5 \cdot 10^{-10} \text{ m} \cdot \text{rad}$$

and the beam size (1*σ*) is

$$\boldsymbol{\sigma = \sqrt{\varepsilon \cdot \beta} \cdot \sqrt{180 \text{ m} \cdot 5 \cdot 10^{-10} \text{ m} \cdot \text{rad}} = 0.3 \text{ mm}}$$

Figure 14 shows the result of a measurement of the transverse beam size. The points represent the measurement values, the curve a Gaussian fit, to obtain the sigma of the distribution as a number to qualify the beam dimension. In general, we will define the aperture dimensions of the vacuum chamber as a certain multiple of this beam sigma. In the case of the LHC, for example, we require a minimum aperture of $r_0 = 18\sigma$ inside the mini-beta quadrupoles.

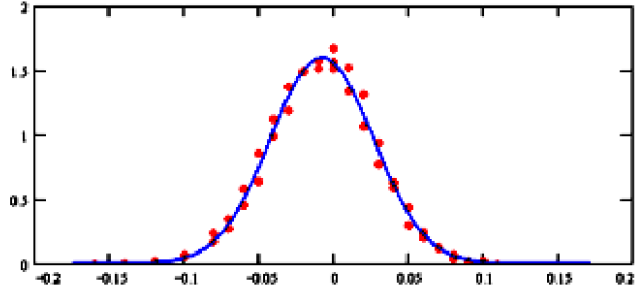

**Fig. 14**: Measured transverse particle density in a beam and a Gaussian fit to obtain the σ of the distribution

## 5   Liouville's theorem during acceleration

There is a special issue concerning the phase space area of a particle beam as soon as we start to accelerate. It has been explained that the expression of the beam emittances

$$\varepsilon = \gamma(s)\,x^2(s) + 2\alpha(s)x(s)x'(s) + \beta(s)\,x'^2(s)$$

represents an ellipse in the $x$–$x'$ space, which we call in a rather sloppy manner 'phase space'.

The real phase space, however, as defined in classical mechanics, relates the position $x$ with its canonical conjugate momentum $p_x$. It is the area in this real phase space that is conserved according to Liouville's theorem. So if

$$\int p_x\, dx = const$$

and rewriting for the angle

$$x' = \frac{dx}{ds} = \frac{dx}{dt}\frac{dt}{ds} = \frac{\beta_x}{\beta}$$

which is the ratio of the relativistic $\beta$-parameter between the transverse motion and the longitudinal one, we get from Liouville's theorem

$$\int p\, dq = mc\gamma\beta \underbrace{\int x'\, dx}_{\varepsilon} = const$$

with the expression in the integral being our 'phase space'. Still, it is true that the beam emittance is constant for a given energy, but as soon as we start to accelerate $\varepsilon$ will shrink as

$$\varepsilon = \int x'\, dx \propto \frac{1}{\beta\gamma}$$

where $\gamma$ and $\beta$ are the relativistic parameters and should not be confused with the beam optics functions.

As a consequence, it turns out that a proton machine or an electron linac will need the highest aperture at injection energy. As soon as we start to accelerate, the beam size shrinks as $\gamma^{-1/2}$ in both transverse planes. At lowest energy the machine will have major aperture problems, and here we have to minimize the beta function to obtain a beam envelope that fits into the vacuum chamber. At high energy this aperture restriction is more relaxed and we can develop optics solutions that allow for higher values of $\beta$. For example, a mini-beta concept that, as we will see, leads to extreme beta values in the focusing magnets will only be adequate at flat top. An example is shown in Fig. 15 for the case of the HERA proton storage ring. The particles were injected into this machine at an energy of 40 GeV and after acceleration reached a flat-top energy of 920 GeV. For the same beam optics, the figure compares the situation at injection energy and at flat top. Owing to the factor of 23 in energy increase, the beam size shrinks by nearly a factor of 5 in both planes.

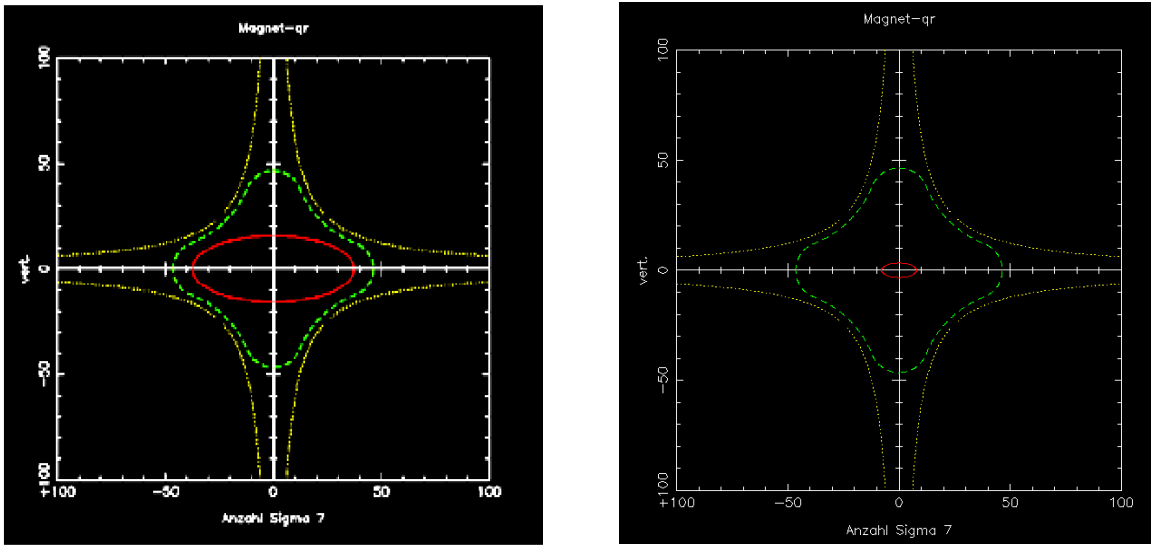

**Fig. 15**: Transverse beam envelope for design parameters of the HERA proton ring at injection and flat top. For the same optics, the beam size shrinks considerably during the acceleration.

In the case of the LHC, the two different beam optics optimized following these needs are shown in Fig. 16. While at injection energy of 450 GeV the maximum beta values have to be limited to 500 m, much higher values are possible at high energy, and close to the mini-beta insertions a maximum of $\beta = 4.5$ km is possible.

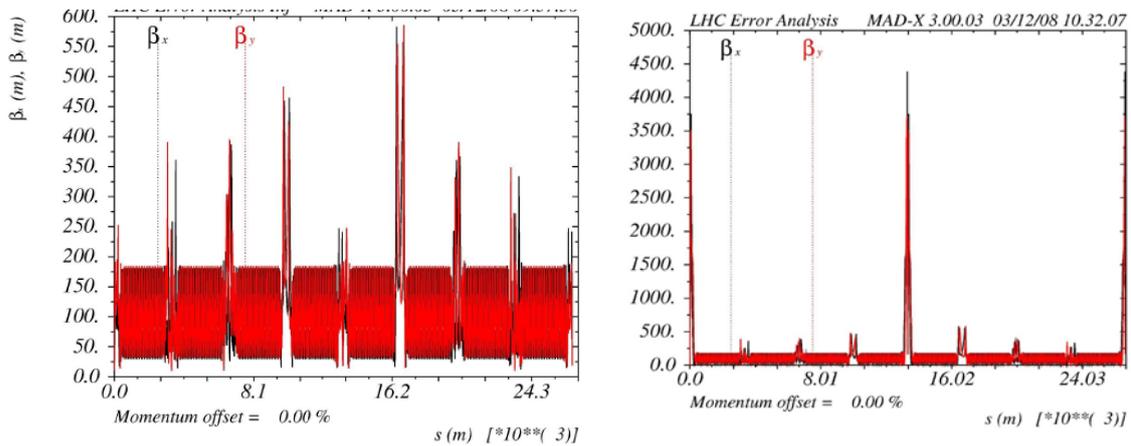

**Fig. 16**: LHC beam optics optimized for injection (left) and luminosity (right). The large beta functions needed for luminosity operation can only be applied at flat-top energy where the beam emittance is considerably reduced.

## 6  Dispersive effects

Until now we have treated the beam and the equation of motion as a mono-energetic problem. Unfortunately, in the case of a realistic beam, we have to deal with a considerable distribution of the particles in energy or momentum. Typical values are

$$\frac{\Delta p}{p} \approx 1.0\ 10^{-3}$$

This momentum spread will lead to several effects concerning the bending of the dipole magnets and the focusing strength of the quadrupoles. It turns out that the equation of motion, which was a homogeneous differential equation until now, will get a non-vanishing term on the right-hand side:

$$x'' + x(\frac{1}{\rho^2} - k) = \frac{\Delta p}{p} \cdot \frac{1}{\rho}$$

The general solution therefore is the sum of the solution of the homogenous equation of motion plus a special solution of the inhomogeneous one:

$$x(s) = x_h(s) + x_i(s)$$

Here $x_h$ is the solution that we have discussed until now and $x_i$ is an additional contribution that has yet to be determined. For convenience, we usually normalize this second term and define a function, the so-called dispersion:

$$D(s) = \frac{x_i(s)}{\Delta p / p}$$

This describes the dependence of the additional amplitude of the transverse oscillation on the momentum error of the particle. In other words it fulfils the condition

$$x_i''(s) + K(s) \cdot x_i(s) = \frac{1}{\rho} \cdot \frac{\Delta p}{p}$$

The dispersion function is usually calculated by optics programs in the context of the calculation of the usual optical parameters and is of equal importance. Analytically it can be determined for single elements via

$$D(s) = S(s) \int_{s0}^{s1} \frac{1}{\rho} C(\tilde{s}) d\tilde{s} - C(s) \int_{s0}^{s1} \frac{1}{\rho} S(\tilde{s}) d\tilde{s}$$

where $S(s)$ and $C(s)$ correspond to the sine-like and cosine-like elements of the single-element matrices in the lattice [6].

Typical values in the case of a high-energy storage ring are

$$x_\beta = 1\text{–}2 \text{ mm}, D(s) = 1\text{–}2 \text{ m}$$

and, for a typical momentum spread of $\Delta p / p \approx 10^{-3}$, we obtain an additional contribution to the beam size from the dispersion function that is of the same order as the one from the betatron oscillations $x_\beta$. An example of a high-energy beam optics including the dispersion function is shown in Fig. 17.

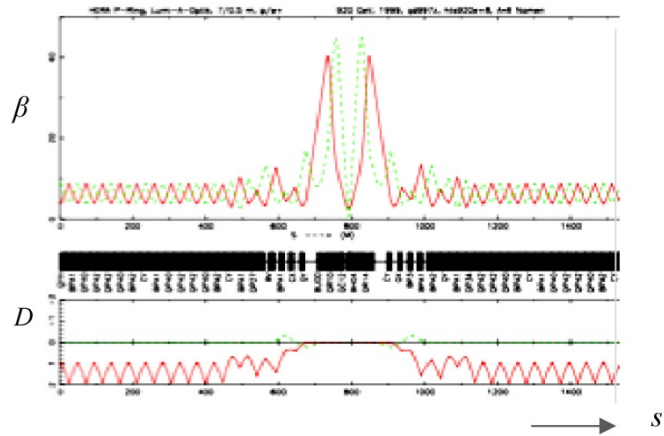

**Fig. 17**: Beam optics function β(s) (upper part) and dispersion function D(s) (lower part) for a part of a typical collider ring. The dispersion must vanish at the collision point in the middle of the horizontal axis.

It should be pointed out that the dispersion describes that special orbit that an ideal particle would have in the absence of no betatron oscillations ($x_\beta = x_\beta' = 0$) for a momentum deviation of $\Delta p/p = 1$. Still, it describes 'just another particle orbit' and so it is subject to the focusing forces of the lattice elements, as seen in Fig. 17.

## 7  Mini-beta insertions and luminosity

The straight sections of a storage ring are often designed for the collision of two counter-rotating beams. For a given overall cross-section of the particle collision (i.e., the probability for a physics process to occur during the interaction of the particles), the event rate of the collision process is determined by the so-called luminosity of the storage ring.

$$R = \sigma_R \cdot L$$

Schematically the situation is shown in Fig. 18.

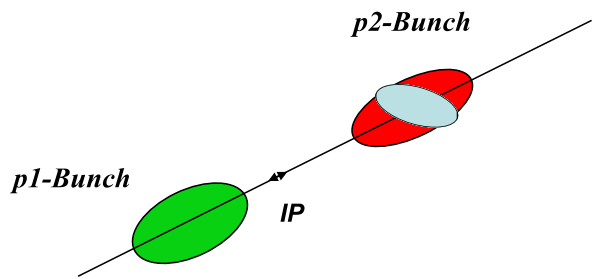

**Fig. 18:** Colliding bunches and luminosity

The luminosity is characterized by the storage ring parameters and depends on the stored beam current and the transverse size of the colliding bunches at the interaction point:

$$L = \frac{1}{4\pi e^2 f_0 n_b} * \frac{I_{p1} I_{p2}}{\sigma_x \sigma_y}$$

Here $I_{p1}$ and $I_{p2}$ are the values of the stored beam currents, $f_0$ is the revolution frequency of the machine, and $n_b$ is the number of stored bunches. The quantities $\sigma_x^*$ and $\sigma_y^*$ in the denominator are the beam sizes in the horizontal and vertical planes at the interaction point. For a high-luminosity collider, the stored beam currents have to be large and, at the same time, the beams have to be focused at the interaction point to very small dimensions. The β-functions at the collision points are therefore very small compared with their values in the arc cells. Typical values are more in the range of centimetres than of metres. At the same time, a large drift space is needed where focusing elements cannot be installed due to the particle detector that will be placed around the interaction point (Fig. 19).

As an example, the parameter list of the LHC including the design luminosity is presented:

$$\beta^*_{x,y} = 0.55 \text{ m} \qquad f_0 = 11.245 \text{ kHz}$$
$$\varepsilon_{x,y} = 5 \cdot 10^{-10} \text{ m} \cdot \text{rad} \qquad n_b = 2808 \qquad \Bigg\} \quad L = 1.0 \cdot 10^{34} \text{ cm}^{-2}\text{s}^{-1}$$
$$\sigma_{x,y} = 17 \text{ μm} \qquad I_p = 584 \text{ mA}$$

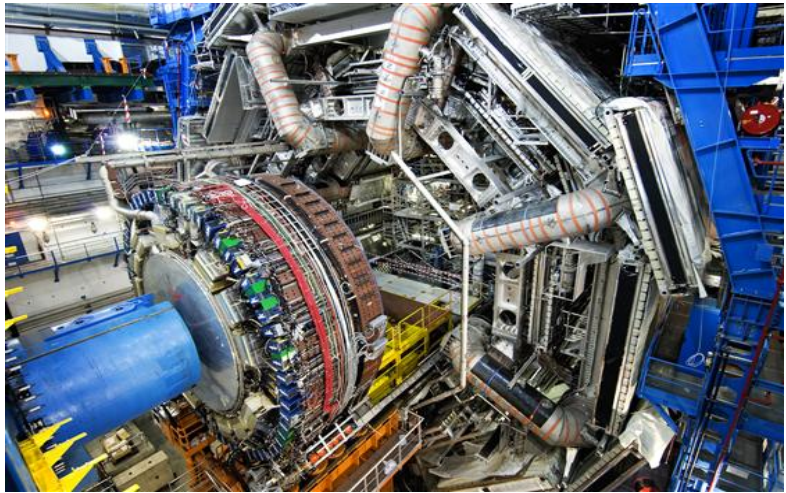

**Fig. 19:** Particle detector of the ATLAS collaboration at the LHC storage ring

Figure 20 shows the typical layout of such a mini-beta insertion. It consists in general of:
- a symmetric drift space that is large enough to house the particle detector and whose beam waist (where $\alpha_0 = 0$) is centred at the interaction point (IP) of the colliding beams;
- a quadrupole doublet (or triplet) installed on each side as close as possible to the IP;
- additional quadrupole lenses to match the optical parameters of the mini-beta insertion to the optical parameters of the lattice cell in the arc.

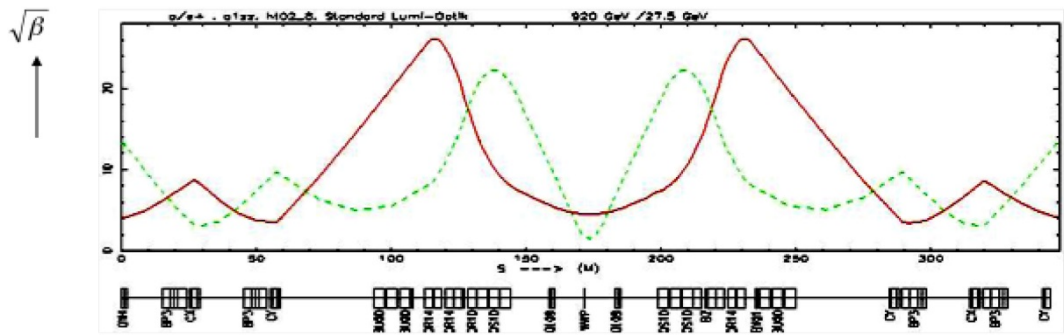

**Fig. 20:** Layout of a mini-beta insertion scheme

Following Liouville's theorem, the request of a small beam size immediately will lead to a corresponding increase of beam divergence. Figure 21 shows the corresponding situation in phase space. It compares the *x*–*x'* ellipse in a typical location of the storage ring with the situation obtained at the IP.

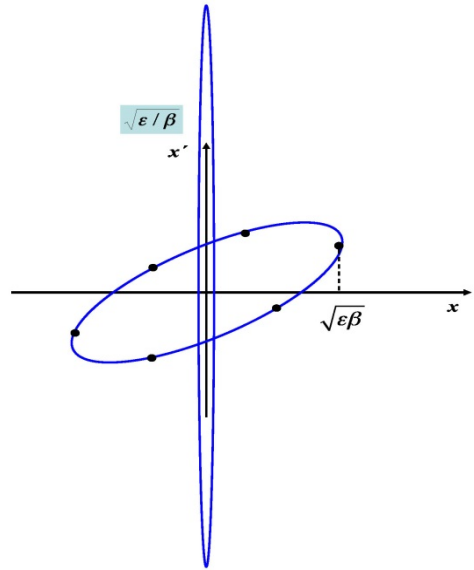

**Fig. 21:** Phase space ellipse in the arc of a storage ring and at the interaction point

While the beam size shrinks with reduced $\beta$ as $\sigma = \sqrt{\varepsilon\beta}$, the corresponding divergence is increased as $\sigma' = \sqrt{\varepsilon/\beta}$. More quantitatively, the behaviour of the beta function in the vicinity the IP can be calculated and we obtain within the symmetric drift space of the mini-beta insertion

$$\beta(s) = \beta_0 + \frac{\ell_1^2}{\beta_0}.$$

Inevitably, the strong focusing that is needed to obtain the smallest spot size of the beam leads to a large increase of the beam inside the first quadrupoles after the IP. It is at this location that we will need the strongest and at the same time the largest aperture quadrupoles – this is the price that we have to pay for the luminosity that we will get out of the design.

**References and further reading**